\documentclass[12pt]{article}
\usepackage{epsf,epsfig,amsmath}
\begin{document}

\title{Classical information and Mermin's non-technical proof
of the theorem of Bell}

\author{Karl Hess$^1$ and Walter Philipp$^2$}

\date{$^1$ Beckman Institute, Department of Electrical Engineering
and Department of Physics, University of Illinois, Urbana, Il
61801
\\ $^{2}$ Beckman Institute, Department of Statistics and Department of
Mathematics, University of Illinois,Urbana, Il 61801 \\ }
\maketitle

\begin{abstract}
We show that Mermin's reasoning against our refutation of his
non-technical proof for Bell-type inequalities is of limited
significance or contains mathematical inconsistencies that, when
taken into account, do not permit his proof to go forward. Our
refutation therefore stands.

\end{abstract}

We discuss two recent notes of Mermin \cite{recmer},
\cite{recmer1} that deal with our refutation \cite{hp} of Mermin's
non-technical proof \cite{mermin} of Bell's inequalities and other
publications related to this discussion \cite{hpp1}-\cite{gref}.
We first show, that the latest note \cite{recmer1} does not add
any substance to our original refutation \cite{hp}, \cite{comm}.
Secondly, we give a more detailed explanation of our original
refutation.

Mermin discusses in his latest note \cite{recmer1} the
admissibility of certain classical information that can be
exchanged between the stations $S_1$ and $S_2$ in Einstein-
Podolsky-Rosen (EPR) type experiments. All classical information
is permitted except `` ...any information whatever about the
setting it has randomly been given in that run." Mermin writes:
``Let us turn Hess and Philipp upside down and explore the extent
to which Bell's theorem survives, not only if, following Hess and
Philipp, we take advantage of properties of the detectors
correlated by the time on local synchronized clock's, but even if
we allow further correlation of the detectors through direct
straightforward ongoing classical communication between them."
What Mermin does not appreciate here is, that our introduction of
time as an independent variable in Bell's functions $A, B$ adds
not only time as a variable but also adds the set of all functions
of time and settings. Indeed, we have stated repeatedly and
clearly that our extension of Bell's parameter space is the
addition of functions of time and settings, namely the addition of
$\bf {time}$ and $\bf {setting}$ dependent parameter random
variables ${\lambda}_{{\bf a},t}^*$, ${\lambda}_{{\bf b},t}^*$,
${\lambda}_{{\bf c},t}^*$ for station $S_1$ and ${\lambda}_{{\bf
a},t}^{**}$, ${\lambda}_{{\bf b},t}^{**}$, ${\lambda}_{{\bf
c},t}^{**}$ for station $S_2$. The structure of these functions
may be constituted such as to carry information about some of the
history of how and when the settings were and are actually chosen
and thus may contain information on the setting of the given run.
Therefore these functions (parameter random variables) cannot be
communicated during the course of each run to the other station.
Mermin's addition of classical information is then of very limited
significance to our enlargement of the parameter space and we are
back to the previous discussion whether our addition is sufficient
to refute Mermin's model in the first place. We add here further
explanations to this discussion which should help to better
understand our initial refutation of Mermin's proof \cite{recmer},
\cite{comm}.

To facilitate the discussion, we attempt to provide a one to one
correspondence of Mermin's notation and ours. We cannot complete
our reasoning with Mermin's all too abbreviated non-technical
notation and therefore need to take this step. We also proceed now
in two stages. We first consider only the parameters \cite{mermin}
of the original publication of Mermin which are all parameters
``that in each run both particles carry to their detectors" and
only later discuss the enlarged set of parameters. We therefore
consider random variables $A=\pm1$ in station $S_1$ and $B=\pm1$
in station $S_2$ that describe the potential outcome of spin
measurements and are indexed by instrument settings that are
characterized by three-dimensional unit vectors ${\bf a}, {\bf b},
{\bf c}$ in both stations. Mermin introduces no precise
counterpart for the values that these random variables $A, B$ may
assume as distinguished from the actual outcomes in form of green
and red detector flashes (the data). These green and red flashes
correspond then also to the values that $A, B$ may assume: $+1$
which we can identify with green and $-1$ which we can identify
with red.

The key assumption of Bell \cite{bellbook} and also of Mermin in
his original non-technical proof \cite{mermin} is that the random
variables $A, B$ depend only on the setting in the respective
station and on another random variable $\Lambda$ that carries the
information with the particles that are emitted from a common
source. Because of Einstein locality and the particular way EPR
experiments are performed \cite{eprex}, the parameter random
variables $\Lambda$ are independent of the settings. Mermin uses
instead of $\Lambda$ instruction sets e.g. GGR meaning flash green
for settings ${\bf a}, {\bf b}$ (which Mermin actually labels $1,
2$) and flash red for setting ${\bf c}$ (labelled $3$ by Mermin).
In our and Bell's notation this means that for a particular value
$\Lambda^*$ that the variable $\Lambda$ may assume and that
corresponds to the specific instruction set GGR we have $A({\bf
a}, \Lambda^*) = A({\bf b}, \Lambda^*) = +1$ and $A({\bf c},
\Lambda^*) = -1$. The following Table \ref{TA:ma} summarizes the
eight possible instruction sets and the nine possible different
$AB$ products which are used by Mermin as a model for
EPR-experiments.
\begin{table}[ht]
  \begin{center}
    \begin{tabular}{|l||r|r|r||r|r|r||r|r|r|}\hline
      ${\Lambda}$ &
${A_{\bf a}}{B_{\bf a}}$ & ${A_{\bf a}}{B_{\bf b}}$ & ${A_{\bf
a}}{B_{\bf c}}$ & ${A_{\bf b}}{B_{\bf a}}$ & ${A_{\bf b}}{B_{\bf
b}}$ & ${A_{\bf b}}{B_{\bf c}}$ & ${A_{\bf c}}{B_{\bf a}}$ &
${A_{\bf c}}{B_{\bf b}}$ & ${A_{\bf c}}{B_{\bf c}}$ \\ \hline
      RRR & $+1$ & $+1$ & $+1$ & $+1$ & $+1$ & $+1$ & $+1$ & $+1$ & $+1$\\ \hline
      RRG & $+1$ & $+1$ & $-1$ & $+1$ & $+1$ & $-1$ & $-1$ & $-1$ & $+1$\\ \hline
      RGR & $+1$ & $-1$ & $+1$ & $-1$ & $+1$ & $-1$ & $+1$ & $-1$ & $+1$\\ \hline
      GRR & $+1$ & $-1$ & $-1$ & $-1$ & $+1$ & $+1$ & $-1$ & $+1$ & $+1$\\ \hline
      GGR & $+1$ & $+1$ & $-1$ & $+1$ & $+1$ & $-1$ & $-1$ & $-1$ & $+1$\\ \hline
      GRG & $+1$ & $-1$ & $+1$ & $-1$ & $+1$ & $-1$ & $+1$ & $-1$ & $+1$\\ \hline
      RGG & $+1$ & $-1$ & $-1$ & $-1$ & $+1$ & $+1$ & $-1$ & $+1$ & $+1$\\ \hline
      GGG & $+1$ & $+1$ & $+1$ & $+1$ & $+1$ & $+1$ & $+1$ & $+1$ & $+1$\\ \hline
   \end{tabular}
   \caption{Possible (but exclussive) $AB$ products}\label{TA:ma}
  \end{center}
\end{table}
According to Mermin's point (i) \cite{recmer1}, the columns of
${A_{\bf a}}{B_{\bf a}}$, ${A_{\bf b}}{B_{\bf b}}$ and ${A_{\bf
c}}{B_{\bf c}}$ all have entries $+1$. It is clear, and this is
the main point of Mermin's argument, that each of the 8 rows of
the Table \ref{TA:ma} contains at least five entries $+1$ and at
most 4 entries $-1$. Each pair of settings occurs with probability
$1/9$. Denote by $p_1,...,p_8$ the probabilities that the
instruction set $\Lambda$ of rows $1,...,8$ is carried by both
particles in a given run. Then, no matter of whether or not we
consider actual outcomes or potential outcomes, the average over
all possibilities obeys
\begin{equation}
{\frac {1} {9}}[9p_1 + 9p_8+ (5-4)(p_2 + p_3 + p_4 + p_5 + p_6 +
p_7)] \geq {\frac {1} {9}} \label{mr21}
\end{equation}
as ${\sum}_{l=1}^8 p_l = 1$, thus contradicting point (ii) of
references \cite{recmer}, \cite{recmer1}. This represents Mermin's
non-technical argument in a more formal way. As can easily be seen
from Table \ref{TA:ma}, it is crucial that $A_{\bf a}$ in Table
\ref{TA:ma} is the same in each of the three entries where it
appears and that $A_{\bf a} = B_{\bf a}$ (and similarly for
$A_{\bf b}$ etc.).

We now turn to the expanded instruction set including time and
setting dependent instrument parameters. We emphasize that the
time dependence is crucial and all our reasoning rests on it. If
it could be shown that EPR experiments contain no time
dependencies and all elements of a corresponding model are also
independent of time, then we have no argument. Let us therefore
assume that the experiments contain time dependencies and that the
theory needs to include time and setting dependent parameter
random variables. It becomes then a question of which mathematical
objects correspond to Mermin's instruction sets. Following Mermin
we identify now the instruction sets with instructions or
operations that turn $A$ into $\pm 1$ and $B$ into $\pm 1$, i.e.
in Mermin's newest definition the instruction sets are the random
variables $A$ and $B$ \cite{foot1}. We can now again consider
Table \ref{TA:ma} with the same notation except that the
instruction sets refer to the expanded instruction sets. However,
we have lost now the clear distinction that Einstein locality
necessitates for source and instrument parameters, respectively.
More importantly, it can no longer be guaranteed that $A_{\bf a}$
in Table \ref{TA:ma} is the same in each of the three entries (and
similar for $A_{\bf b}$ etc.). Furthermore, we have $A_{\bf a} =
B_{\bf a}$ only for the same time.  This was explained by us in
Eqs.(1) and (2) of reference \cite{comm}. However, this point
contains some subtle reasoning and we give therefore the following
elaboration. For a theory to provide valid conclusions, special
care must be taken when combining, adding and/or counting
different elements (actual or imagined) that are mutually
exclusive. In the case that we consider there are two further
subtleties. First, different elements of the theory are mutually
exclusive only at the same time and second, the measurement time
itself appears as a random variable. This puts restrictions on the
possible measurement times which now must be different for
different settings. If they are chosen to be equal, Table
\ref{TA:ma} will contain mutually exclusive alternatives that
cannot simultaneously be used in mathematical operations such as
counting, adding or averaging. We will return to this point below.

The instruction sets become now also time and setting dependent
and the joint frequency of occurrence of the setting dependent
instrument parameters may now be different in different columns of
Table \ref{TA:ma}. We see that Mermin's proof, that is based on
the equality of all $A_{\bf a}$, $A_{\bf b}$ and $A_{\bf c}$ (and
similarly for the $B$'s) in Table \ref{TA:ma}, comes to a halt.
His argument \cite{recmer} that the expanded instruction sets for
a given single experiment in one time interval must be one of his
eight instruction sets is true but meaningless. It has no
consequences for the statistics and the associated probability
measure and therefore does not guarantee Mermin's way of counting
$+1$ or $-1$.

One way one still could attempt to proceed with the proof is to
assume that EPR experiments are equivalent to other experiments
that are made all at the same time. This does not only involve
counterfactual reasoning but contradicts the assumed fact of
possible time dependencies. The only other way to proceed with
Mermin's proof is to add to the possible outcomes of the Table
\ref{TA:ma} the outcomes for all the nine settings as they would
have been obtained when taken at the same time and to include them
into the counting of positive and negative $AB$ products. This
procedure, however, again does not only contain counterfactual
reasoning, but contradicts the facts because then the theory $\bf
{counts}$ also impossible outcomes by adding mutually exclusive
alternatives. This, in turn, has the consequence that the set of
elements so counted is nine times larger than the set of
experimental results. Any reasonable procedure to establish a
theoretical model for an experiment will require a one to one
correspondence of what is added/counted/averaged in theory to what
is added/counted/averaged in the experiment. We conclude that
Mermin's proof, as well as those of other Bell type inequalities,
apply only to a stationary situation that may have nothing to do
with past and present EPR experiments \cite{eprex}.

We finish this discussion with an example of what can happen if
imagined alternatives are added (or counted) that involve,
simultaneously (time!), mutually exclusive alternatives. Mutually
exclusive events are discussed in every probability text well
before the concept of countable additivity of probability measures
is introduced. When probabilities of mutually exclusive events are
added, it is understood that the simultaneous occurrence of
mutually exclusive events is impossible. Our example below has
nothing to do with quantum mechanics. It just illustrates the
rules of standard probability theory that must be obeyed. Consider
a coin with a little magnet inside and a hidden bigger magnet
under ground. The experimenter can set the underground magnet N or
S and correspondingly the coin will show with higher probability
head or tail, respectively, in any given experiment. If one
performs a large number of coin tosses, the result will be biased
according to the choices that are made for the underground magnet
(N, S). Now in analogy to the above addition of nine terms for
each single run of the experiment, construct a theory in which one
simply adds or counts, simultaneously, the mutually exclusive
alternatives of head or tail. Count 1 if a head shows and count 0
if a tail shows in each of the potential outcomes. Then for each
toss the count of potential heads plus the count of potential
tails equals 1. Because we have twice as many potential outcomes
than actual tosses, the likelihood of heads equals exactly $1/2$.
Hence such a theory necessarily concludes that this is a fair
game. This theory, however, has combined simultaneously mutually
exclusive alternatives. It necessarily counts twice as many
elements than any given set of experimental results contains and
is therefore not admissible. This is, of course, immediately
obvious to anyone.

As mentioned above, the reasoning becomes more subtle when time
plays a role in the random variables and the elements become
mutually exclusive at the same time, hence impossible. Then care
must be exercised in the possible choices of the random variables
related to measurement time or the same mistake as outlined in the
above example could be made. One way to proceed safely is to label
the measurement times by the actual setting that is chosen e.g.
$t_{{\bf a}{\bf b}}$, $t_{{\bf a}{\bf c}}$ etc. and regard
measurement times with different label as different random
variables. Such labelling clearly shows the special role of
measurement time as random variable.

We present now our argument in a still more technical way to bring
the problems of Mermin's reasoning into focus. Consider the random
variables $A_{\bf i} = A_{\bf i}(\Lambda)$ and $B_{\bf j} = B_{\bf
j}(\Lambda)$ with ${\bf i},{\bf j} = {\bf a}, {\bf b}, {\bf c}$
and which, just as above, assume the values $\pm 1$. Let $\mu$ be
the probability measure that assigns weight $p_l$ to the $\Lambda$
in the $l$'th row. Then according to Mermin's (i) and (ii)  we
have

\begin{itemize}

\item[(i)]

\begin{equation}
A_{\bf j}(\Lambda) = B_{\bf j}(\Lambda) \text{   ,   } {\bf j} =
{\bf a}, {\bf b}, {\bf c} \nonumber
\end{equation}
and

\item[(ii)] the average over all products $A_{\bf i}B_{\bf j}$ is
about 0.

\end{itemize}

The average for actual as well as potential outcomes equals about:
\begin{equation}
{\frac{1} {9}} \int ({A_{\bf a}(\Lambda)} + {A_{\bf b}(\Lambda)} +
{A_{\bf c}(\Lambda)})({B_{\bf a}(\Lambda)} + {B_{\bf b}(\Lambda)}
+ {B_{\bf c}(\Lambda)})d\mu \label{mr22}
\end{equation}
If we now consider the original instruction sets that are
independent of the setting then we have by Mermin's condition (i)
\begin{equation}
{A_{\bf a}(\Lambda)} + {A_{\bf b}(\Lambda)} + {A_{\bf c}(\Lambda)}
= {B_{\bf a}(\Lambda)} + {B_{\bf b}(\Lambda)} + {B_{\bf
c}(\Lambda)} \label{mr23}
\end{equation}
for any given value of $\Lambda$ (which could be e.g.
$\Lambda^*$). This gives for Eq.(\ref{mr22})
\begin{equation}
{\frac{1} {9}} \int {({A_{\bf a}(\Lambda)} + {A_{\bf b}(\Lambda)}
+ {A_{\bf c}(\Lambda)})}^2d\mu \geq {\frac{1} {9}} \label{mr24}
\end{equation}
contradicting Mermin's condition (ii).

However, if we use now the enlarged instruction set which includes
both time and setting dependent parameters, then the $AB$ products
that involve different pairs of settings involve different times
and the parameter random variables $\Lambda$ in Eq.(\ref{mr23})
may all be different. In fact, just as above, it can no longer be
guaranteed that $A_{\bf a}$ is the same in each of the three
products $A_{\bf a}B_{\bf a}$, $A_{\bf a}B_{\bf b}$ and $A_{\bf
a}B_{\bf c}$ and that $A_{\bf a} = B_{\bf a}$, independent of
time. Again, Mermin's proof comes to halt.

One final comment. It is obvious that the above argument still
works for an arbitrary probability measure $\mu$, that is for any
probability distribution governing $\Lambda$ of the original
instruction set \cite{mermin}. However, if the enlarged
instruction set is considered, Mermin's proof cannot be completed
unless it can be guaranteed that $A_{\bf a}$ (or at least its
probability distribution) is the same in each of the three
products $A_{\bf a}B_{\bf a}$, $A_{\bf a}B_{\bf b}$ and $A_{\bf
a}B_{\bf c}$. More specifically one needs to take into account
that there are $\bf {nine}$ joint distributions in operation
namely $\mu_{{\bf a}{\bf a}}$, $\mu_{{\bf a}{\bf b}}$, $\mu_{{\bf
a}{\bf c}}$,..., $\mu_{{\bf c}{\bf c}}$. Thus, unless it can be
guaranteed that the first marginal distribution of the three
distributions $\mu_{{\bf a}{\bf a}}$, $\mu_{{\bf a}{\bf b}}$ and
$\mu_{{\bf a}{\bf c}}$ is the same, the proof cannot go forward.

We conclude that our refutation of Mermin's non-technical proof
for the theorem of Bell stands.

Acknowledgement: the work was supported by the Office of Naval
Research N00014-98-1-0604, the MURI supported by ONR and by NSF
Grant CCR-0121616. We thank  Salvador Barraza-Lopez for providing
sources regarding language, valuable discussions and help with the
manuscript and Istvan Berkes for valuable discussions.

\end{document}